\documentclass[11pt]{article}

\usepackage{amsthm,amsmath,amssymb,bm}
\usepackage{graphicx,color,epsfig}
\usepackage[margin=1in]{geometry}

\newcommand{\N}{\mathbb{N}}
\newcommand{\R}{\mathbb{R}}

\newtheorem{Thm}{Theorem}[section]
\newtheorem{Claim}[Thm]{Claim}
\newtheorem{Pro}[Thm]{Proposition}
\newtheorem{Lem}[Thm]{Lemma}

\begin{document}

\title{Convergence to stationary points in the Weisbuch-Kirman-Herreiner model for buyers' preferences in fish markets}
\author{Ali Ellouze and Bastien Fernandez}
\date{}
\maketitle

\begin{center}
Laboratoire de Probabilit\'es, Statistique et Mod\'elisation\\
CNRS - Univ. Paris Cit\'e -  Sorbonne Univ.\\
Paris, France\\
ellouze@lpsm.paris and fernandez@lpsm.paris
\end{center}

\begin{abstract} 
In a paper published in The Economic Journal in 2000, Weisbuch et al.\ introduce a model for buyers' preferences to the various sellers in over-the-counter (OTC) fish markets. While this model has become an archetype of economic conceptualization that combines bounded rationality and myopic reasoning,  the literature on its asymptotic behaviours has remained scarce. In this paper, we proceed to a mathematical analysis of the dynamics and its full characterization in the simplest case of homogeneous buyer populations. By using elements of the theory of cooperative dynamical systems, we prove that, independently of the number of sellers and parameters, for almost every initial condition, the subsequent trajectory must asymptotically approach a stationary state. Moreover, for simple enough distributions of the sellers' attractiveness, we determine all stationary states and their parameter-dependent stability. This analysis shows that in most cases, the asymptotic preferences are ordered as the attractiveness are. However, depending on the parameters, there also exist robust functioning modes in which those sellers with highest preference are not the ones that provide highest profit. 
\end{abstract}

\section{Introduction and definitions}
The paper \cite{WKH00} aims at capturing in a nutshell the main drives of buyer behaviours in OTC wholesale fresh product markets. Following behavioural approaches to economic theory  \cite{BD01,BH97,G12} -- which in the current case, consists in investigating how trading interpersonal relationships and imperfect information impact market functioning --  based on empirical evidences from the Marseille professional fish market, and inspired from agent-based models in Statistical Physics \cite{FF09}, a simple stylized model is introduced and discussed. For markets composed by 2 sellers and assuming homogeneous buyer populations, analytical results are given on the existence and stability of stationary points and their bifurcations. Insights from numerics are also provided in more elaborated cases.

While that paper has become a landmark in the economic literature, to the best of our knowledge, surprisingly, no complete or even systematic analysis of the dynamics has been published, even in the simple case of homogeneous buyers. Aiming at filling this gap, the present paper characterizes the asymptotic dynamics at large times of the model in \cite{WKH00} for homogeneous-buyers markets with arbitrary number of sellers, and for arbitrary values of the parameters. In particular, we prove that almost every trajectory converges to a stationary point, in line with basics of neo-classical economy. 
The proof consists in showing that a number of constraints applies to the trajectories, which implies that the dynamics is eventually controlled by some (trajectory-dependent) cooperative and irreducible system \cite{H85}, from where the conclusion follows suit. In order to illustrate that phenomenology, we also compute all stationary states and their parameter-dependent stability and bifurcations, for simple distributions of the sellers' attractiveness. 

The paper is organised as follows. After the definition and basic considerations on parameters, we state in Section 2 the main result about convergence to stationary points. For certain instances of parameters, one can actually prove that all trajectories converge to a stationary point. All these results are proved in Section \ref{S-ANALYSIS}. In Section \ref{S-STATPT}, we determine the existence and the stability of all stationary points in illustrative instances of parameters. The simplest case is when all sellers have equal attractiveness. We also consider the case of a market of 2 sellers as in \cite{WKH00}, and the more general case of two-clusters dynamics, in which, and unlike in the case of 2 sellers, the friction-coefficient dependence of the number and stability of stationary points turns out to be non-monotonous. The paper concludes by some considerations about perspectives and future works. 
 
\paragraph{Definition of the dynamics.} 
Let the arbitrary integer $N\in\{2,3,4,\cdots\}$ be the number of sellers in the market (NB: We do not consider the trivial case $N=1$). Assuming that the buyers' population is homogeneous, the model introduced in \cite{WKH00} reduces to a system of coupled ordinary differential equations for the variable $\mathbf{J} = \{J_1,\cdots,J_N\}$, where $J_i\in\R_\ast^+$ represents the buyers' preference toward seller $i$. More specifically, the system that governs the time evolution of $t\mapsto \mathbf J(t)$ is given by 
\begin{equation}
\dot{J}_i = F_i(\mathbf{J}),\ \forall i \in [N]=\{1,\cdots,N\} \quad \textrm{where } \quad F_i(\mathbf{J}): =-\gamma J_i + a_i \frac{e^{ J_i}}{\sum_{j=1}^{N} e^{J_j}},
\label{DYN_WKH}
\end{equation}
and the parameters $\gamma$ and $\mathbf a=\{a_i\}_{i=1}^N$, which are arbitrary, respectively in $\R_\ast^+$ and in $(\R_\ast^+)^N$, can be interpreted as follows.
\begin{itemize}
\item The friction coefficient $\gamma \in \R_\ast^+$ quantifies buyers' memory; by contrast, the sensitivity of the variable $J_i$ to external inputs.
\item The sellers' attractiveness $a_i\in \R_\ast^+$ represent the profits that the buyers make in trading with $i$. 
\end{itemize}

\paragraph{Considerations about the parameters}
In \cite{WKH00}, the second term in the expression of $F_i$ contains a (so-called temperature) parameter $\beta > 0$, \textit{i.e.}\ it writes  $a_i\frac{e^{\beta J_i}}{\sum_{j=1}^{N} e^{\beta J_j}} \in (0,1)$. However such a parameter can be absorbed in the variable $J_i$ through the variable change $J_i \mapsto \beta J_i$, and provided that the parameters $\gamma$ and $\mathbf a$ are scaled accordingly. In other words, this means that, for every $\beta > 0$, the dynamics of this system is equivalent to the dynamics of \eqref{DYN_WKH}. 

Along the same lines, notice that, up to a rescaling of time, the friction coefficient $\gamma$ can be absorbed in the attractivenesses $\mathbf a$. That is to say, the function $t \mapsto \mathbf{J}(t)$ is a trajectory of \eqref{DYN_WKH} with parameters $\gamma$ and $\mathbf{a}$ if and only if $t \mapsto \mathbf{J}(\frac {t}{\gamma})$ is a trajectory of \eqref{DYN_WKH} with parameters $\gamma'=1$ and $\mathbf a'=\frac{1}{\gamma}\mathbf{a}$. However, we keep the two parameters separated for the sake of the interpretation. 

Finally, notice also that, as a particular instance of coupled map system with mean-field coupling \cite{CF05}, the dynamics commutes with the simultaneous permutations of the preferences and attractiveness. That is to say, if $t \mapsto \mathbf J(t)$ is a trajectory of \eqref{DYN_WKH} for the attractiveness $\boldsymbol{a}$ then, for every permutation $\pi$ of $[N]$, the map $t \mapsto \pi J(t)$ where 
\[
\pi \mathbf x=(x_{\pi(1)},\cdots,x_{\pi(N)}),
\]
is a trajectory for the attractiveness $\pi \mathbf{a}$. Accordingly, throughout the paper, we assume without loss of generality (w.l.o.g.)\ that the {\bf individual attractiveness are monotonically ordered} namely, we have  $a_1\leq a_2\leq \cdots \leq a_N$. 

\section{Convergence to stationary points}
This section presents the main result of the paper, namely the almost sure convergence to stationary points ({\sl ie.}\ the zeros of $F$). We first ensure that the dynamics is globally well defined, so that it makes sense to investigate the asymptotic behaviour of the trajectories. 
\begin{Thm}
(i) For every initial condition $\mathbf{J}^0 \in \R^N$, the equation \eqref{DYN_WKH} admits a unique solution $t \mapsto \mathbf{J}(t)$ such that $\mathbf{J}(0)=\mathbf{J}^0$, which is defined for all $t \in \R^+_\ast$ and is bounded.

\noindent
(ii) For Lebesgue every initial condition $\mathbf{J}^0 \in \R^N$, the subsequent trajectory $t \mapsto \mathbf{J}(t)$ asymptotically converges, as $t\to\ +\infty$, to a stationary point. 
\label{MAINRES1}
\end{Thm}
The proof is given in the first two subsections of the next Section \ref{S-ANALYSIS} as part of the global analysis of the dynamics. This result, which from a practical economical viewpoint, ensures that the agents' preferences asymptotically stabilise, is not obvious as it may appear. Indeed, the expression of the partial derivatives $\frac{\partial F_i}{\partial J_j}$ in the proof of Lemma \ref{EXIST} indicates that the equation \eqref{DYN_WKH} is a competitive dynamical system in the sense of \cite{H85}. Yet, there are examples of competitive systems in which almost every trajectory do not approach a stationary state, starting with the May-Leonard competition model \cite{ML75}. Hence, the proof first consists in proving that the ordering of the coordinates must eventually be constant and the coordinates differences are appropriately bounded in every trajectory. Based on these features, one then shows that the asymptotic dynamics, which turns out to be contained in an $(N-1)$-simplex, is actually governed by a number of systems, which are all cooperative and irreducible. 

Besides, that indicates that the equation \eqref{DYN_WKH} is a competitive system implies a number of properties about the dynamics, in particular
\begin{itemize}
\item for $N\in\N$ arbitrary, if $J,J'$ are two points of a compact $\omega$-limit set, then there exist $i\neq j \in[N]$ such that (Theorem 3.2, Chap.\ 3 in \cite{S95}) 
\[
(J_i-J'_i)(J_j-J'_j)\leq 0.
\] 
\item for $N\in\N$ arbitrary, the flow on a compact $\omega$-limit set is topologically equivalent to a flow of a Lipschitz vector field in $\R^{N-1}$ (Theorem 3.4, Chap.\ 3 in \cite{S95}).
\end{itemize}

Back to the specific model under consideration, depending on the parameters, it can be ensured that all trajectories actually converge to a stationary point, as the next statement claims.
\begin{Pro}
(i) Assume that $N\in \{2,3\}$ or that $N>3$ and $a_i = a$ for all $i \in [N]$ for some arbitrary $a \in \R_+^\ast$. Then, every trajectory converges to a stationary point. 

\noindent
(ii) Assume that $\gamma>\frac{a_N}2$. Then equation \eqref{DYN_WKH} has a unique stationary point which is globally attracting. 
\label{CONTRACT}
\end{Pro}
The proof is given in subsection \ref{S-CONTRACT}. Statement {\em (i)} for $N\in \{2,3\}$ is an immediate consequence of Theorem 2.7 in \cite{H85} for the dynamics of planar competitive or cooperative systems. The result in the homogeneous case of identical attractiveness is an immediate consequence of the fact that the system \eqref{DYN_WKH} is a gradient system in this case. 

The proof of statement {\em (ii)} is a bit more elaborated and relies  on proving that the flow is contracting \cite{LS98} when $\gamma>\max_i\frac{a_i}2=\frac{a_N}2$. Notice that, when the stationary point is unique, its coordinates must be ordered as the coordinates of $\mathbf a$ are, because, as explained in the proof of Theorem \ref{MAINRES1}, the subset of $\mathbf J$ such that $J_i\leq J_{i+1}$ for all $i\in [N-1]$ is invariant under the dynamics. 
 
As we shall see in Section \ref{S-STATPT} below, depending on $\mathbf{a}$, the condition $\gamma > \frac{a_N}{2}$ for the uniqueness of the stationary point may not be sharp. In particular, when the coordinates of $\mathbf{a}$ do not depend on $i$, the stationary point can be unique even though $\gamma \leq \frac{a_N}{2}$. Yet, uniqueness of the stationary state does not always hold and we shall see that multiple stationary points may exist when $\gamma$ is sufficiently small. 

\section{Analysis of the dynamics}\label{S-ANALYSIS}
\subsection{Well-posedness and a globally attracting set}
The purpose of this section is to prove statement {\em (i)} of Theorem \ref{MAINRES1} and also, as already announced, that the asymptotic dynamics lies in some $(N-1)$-simplex which in particular, serves in the proof of statement {\em (ii)}. For consistency, we start by formulating the results to be proved.
\begin{Lem}
For every initial condition $\mathbf{J}^0 \in \R^N$, the equation \eqref{DYN_WKH} admits a unique solution $t \mapsto \mathbf{J}(t)$ such that $\mathbf{J}(0)=\mathbf{J}^0$, which is defined for all $t \in \R^+_\ast$ and is bounded.

\noindent
Moreover, the simplex $S$, contained in the orthant $(\R_\ast^+)^N$, and defined by
\[
S = \{\mathbf{J} \in (\R_\ast^+)^N\ :\ \sum_{i=1}^{N}\frac{J_i}{a_i} = \frac{1}{\gamma}\},
\]
is invariant and globally attracting under the corresponding flow. 
\label{EXIST}
\end{Lem}
In addition to ensuring statement {\em (i)} of Theorem \ref{MAINRES1}, this property indicates that every stationary point must lies in $S$. This feature will be used extensively below when solving the stationary points equation.
\begin{proof}
The partial derivatives of the map $F$ write
\[
\frac{\partial F_i}{\partial J_j} =\left\{\begin{array}{cc}
 -\gamma + a_i \frac{e^{J_i}\sum_{k \neq i}e^{J_k}}{(\sum_{k=1}^{N}e^{J_k})^2}&\text{if}\ i= j\\
-a_i \frac{e^{J_i+J_j}}{(\sum_{k=1}^{N}e^{J_k})^2} &\text{if}\ i\neq j
\end{array}\right.\quad \forall\ (i,j)\in [N]^2,
\]
hence they all exist and are continuous function of $\mathbf J$. Therefore, we have $F\in C^1(\R^N)$. Moreover, the elementary inequality $\frac{ab}{(a+b)^2}<\frac14$ which holds for all $(a,b) \in (\R_\ast^+)^2$, implies the following one
\[
\max_{(i,j) \in [N]^2} \lvert \frac{\partial F_i}{\partial J_j}\rvert \leq \frac{1}{4} \max_{i \in [N]} a_i + \gamma,
\]
{\sl viz.}\ the derivative of $F$ is bounded on $\R^N$. 

Standard results of the theory of ordinary differential equations imply that for every initial condition $\mathbf{J}^0$ a unique solution $t \mapsto \mathbf{J}(t)$ with $\mathbf{J}(0)=\mathbf{J}^0 $, must exist for a small enough time interval $[0,\epsilon]$. In addition, we have 
\[
F_i(\mathbf{J}) \leq -\gamma J_i+a_i ,\ \forall i \in [N].
\]
By the Gronwall inequality, it follows that we have for every trajectory
\[
J_i(t) \leq J_i(0)e^{-\gamma t}+\frac{a_i}{\gamma}(1-e^{-\gamma t}),\ \forall t\in\R^+,i \in [N],
\]
hence the solution must be bounded as long as it exists. In this case, the theory implies that the solution exists for all positive times. In addition, the inequality on the $F_i$ above and the one on $J_i(t)$ imply that the set
\[
\left\{\mathbf J\in \R^N\ :\ J_i\leq \frac{a_i}{\gamma},\ \forall i\in [N]\right\},
\]
is invariant and attracting.

In addition, the subset 
\[
\left\{\mathbf J\in (\R_\ast^+)^N\ :\ J_i\leq \frac{a_i}{\gamma},\ \forall i\in [N]\right\},
\]
is also invariant and attracting, because the map $F$ is Lipschitz continuous and, for each $i$, we have 
\[
\inf_{\mathbf J\in \R^N\ :\ J_i\leq 0\ \text{and}\ J_j\leq \frac{a_j}{\gamma},\ \forall j\neq i}F_i(\mathbf{J})>0,
\]
see in particular Theorem 4.7 in \cite{BS07}.

Finally, the fact that $S$ is invariant and attracting immediately follows from the equality
\[
\sum_{i=1}^{N} \frac{F_i(\mathbf{J})}{a_i} = 1-\gamma \sum_{i=1}^{N}\frac{J_i}{a_i}.
\]
\end{proof}

\subsection{Proof of statement {\em (ii)} of Theorem \ref{MAINRES1}}
The proof consists of various steps; the first one consists in asserting that in every trajectory, the relative ordering of the coordinates of $\mathbf{J}(t)$ must eventually be constant.
\begin{Claim}
Let $t\mapsto \mathbf{J}(t)$ be an arbitrary solution of \eqref{DYN_WKH}. Then, there exist a permutation $\pi$ of $\{1,\cdots ,N\}$ and $t_0\in\N$ such that
\[
J_{\pi(i)}(t)\leq J_{\pi(i+1)}(t),\ \forall i\in [N-1],\ t\geq t_0.
\] 
\label{EVENTORD}
\end{Claim}
A special case of this ordering, which is also a consequence of the ordering of the coordinates of $\mathbf a$, is the fact the system \eqref{DYN_WKH} is monotone \cite{S95}, that is to say, if the inequality above holds at $t=0$ for $\pi=\mathrm{Id}$, then it is preserved for all $t>0$. The same invariance needs not hold when $\pi\neq\mathrm{Id}$; hence the relevance of the Claim, which is due to the {\bf mean-field type} of interactions in this system.
\begin{proof}
Let $j\in [N]$ be arbitrary and define $\boldsymbol\Delta^j=\{\Delta_k^j\}_{k\in [N]\setminus \{j\}}\in\R^{N-1}$ by $\Delta_k^j=J_k-J_j$. From \eqref{DYN_WKH}, for every $i\neq j$, we have
\begin{equation}
\dot\Delta_i^j=G_i^j(\boldsymbol\Delta^j)\quad\text{where}\quad G_i^j(\boldsymbol\Delta^j)=-\gamma \Delta_i^j+\frac{a_i e^{\Delta_i^j} - a_j}{1+\sum_{k\neq j}e^{\Delta_k^j}}.
\label{DIFFDYN}
\end{equation}
Let now $i<j$. The assumption on the coordinates of $\mathbf a$ imply that the RHS must be positive when $\Delta_i^j=0$. As a consequence, the following alternative must hold in every trajectory
\begin{itemize}
\item either $\Delta_i^j(t)<0$ for all $t\in\R^+$,
\item or there exists $\tau_j^j\in\R^+$ such that $\Delta_i^j(t)\geq 0$ for all $t\geq \tau_i^j$. 
\end{itemize}
In particular, there exists $t_i^j\in\R$ (which is either 0 or $\tau_i^j$), such that the relative ordering of $J_i(t)$ and $J_j(t)$ remains the same for all $t\geq t_i^j$. The Claim then easily follows.
\end{proof}

Not only the ordering of the coordinates of $\mathbf{J}(t)$ must eventually remain the same but the expression of $G_i^j$ forces the differences of these coordinates to certain ranges. To see this, given $j\in [N]$, let the set $S_j$ be defined by 
\[
S_j=\left\{\mathbf{J}\in \R^{N}\ :\ J_i-J_j\leq \log\frac{a_j}{a_i},\ \forall i\in [N]\right\}.
\]
\begin{Claim}
Given an arbitrary solution of \eqref{DYN_WKH}, let $\pi$ and $t_0$ be given by Claim \ref{EVENTORD}. Then, we have 
\[
\mathbf{J}(t)\in S_{\pi(N)},\ \forall t\geq t_0.
\]
\end{Claim}
\begin{proof}
We have already established that 
\[
\Delta_i^{\pi(N)}(t)\leq 0,\ \forall i\in [N], t\geq t_0.
\]
Hence, for those indices $i$ such that $i\leq \pi(N)$, there is nothing to prove since we have $a_i\leq a_{\pi(N)}$. 

Assume then that $\log\frac{a_{\pi(N)}}{a_i}<0$ and consider the equation \eqref{DIFFDYN} with $j=\pi(N)$. By investigating separately the two terms of $G_i^{\pi(N)}(\boldsymbol\Delta^{\pi(N)})$, and since, by Lemma \ref{EXIST}, all differences $\Delta_k^{\pi(N)}$ must remain bounded in every trajectory, it easily results that, when $\Delta_i^{\pi(N)}$ runs over the interval $[\log\frac{a_{\pi(N)}}{a_i},0]$, the quantity $G_i^{\pi(N)}(\boldsymbol\Delta^{\pi(N)})$ must be bounded below by a positive number. 

As a consequence, if $\Delta_i^{\pi(N)}(t_1)\in (\log\frac{a_{\pi(N)}}{a_i},0]$ for some $t\in t_1$, then there must exists $t_2>t_1$ such that $\Delta_i^{\pi(N)}(t_2)>0$. This is incompatible with the conclusion of Claim \ref{EVENTORD}; therefore we must have 
\[
\Delta_i^{\pi(N)}(t)\leq \log\frac{a_{\pi(N)}}{a_i},\ \forall i\in [N], t\geq t_0,
\]
as desired. 
\end{proof}
Let $\Sigma_j$ be the analogue of $S_j$ when expressed in the variable $\boldsymbol\Delta_j$, {\sl ie.}\
\[
\Sigma_j=\left\{\boldsymbol\Delta^j\in \R^{N-1}\ :\ \Delta_i^j\leq \log\frac{a_j}{a_i},\ \forall i\in [N]\setminus\{j\}\right\}.
\]
The equation \eqref{DIFFDYN} actually defines an autonomous flow for the $(N-1)$-dimensional variable $\boldsymbol\Delta^j$, that turns to have a crucial property when in $\Sigma_j$ (see \cite{H85} for the definitions).
\begin{Claim}
Let $j\in [N]$ be arbitrary. When restricted to $\Sigma_j$, the system defined by \eqref{DIFFDYN} is cooperative and irreducible.
\end{Claim}
\begin{proof}
For every $i\neq j$, we have 
\[
\frac{\partial G_i^j}{\partial \Delta_\ell^j}=-e^{\Delta_\ell^j}\frac{a_i e^{\Delta_i^j} - a_j}{(1+\sum_{k\neq j}e^{\Delta_k^j})^2},\ \forall \ell\neq i.
\]
The claimed property is literally the fact that all the RHS here must be positive when $\boldsymbol\Delta^j\in \Sigma_j$.
\end{proof}
Clearly, every set $\Sigma_j$ is p-convex in the sense of \cite{H85}. Let $\Sigma'_j\subset \Sigma_j$ be the subset of points those trajectory remains forever in $\Sigma_j$.  From Lemma \ref{EXIST}, every trajectory of \eqref{DIFFDYN} issued from an initial condition in $\Sigma'_j$ must have a compact closure. Altogether, it follows from Theorem 4.1 in \cite{H85} that, for almost every initial condition in $\Sigma'_j$, the subsequent trajectory must asymptotically converge to a stationary point of \eqref{DIFFDYN}.

Let $\{\phi_j^t\}_{t\in\R^+}$ be the flow associated with the equation \eqref{DIFFDYN} and let $D\phi_j^t$ be its derivative. The Liouville formula for $\mathrm{det}D\phi_j^t(\boldsymbol\Delta^j)$ tells us that for every $t\in\R_\ast^t$, the pre-image $\phi_j^{-t}((\Sigma'_j)^c)$ must also have zero Lebesgue measure. In other words, the set of initial conditions $\boldsymbol\Delta^j\in\R^{N-1}$ those trajectory eventually enters $\Sigma_j$ but which does not converge to a stationary point of \eqref{DIFFDYN}, has zero Lebesgue measure in $\R^{N-1}$.

Recall that the asymptotic dynamics of the system \eqref{DYN_WKH} lies in $S$. Hence, the convergence to a fixed point when in $S_j$ does not depend on the coordinate $J_j$. It follows that the set of initial conditions $\mathbf{J}\in\R^N$ those trajectory eventually enters $S_j$ but which does not converge to a stationary point of \eqref{DIFFDYN}, has zero Lebesgue measure in $\R^N$. The Theorem then follows from the fact that every trajectory must eventually enter one of the $S_j$.

\subsection{Proof of Proposition \ref{CONTRACT}}\label{S-CONTRACT}
\paragraph{Proof of statement {\em (i)}.}
For $N\in \{2,3\}$, this is immediate from Theorem 2.7 in \cite{H85} and Lemma \ref{EXIST} given that
\begin{itemize}
\item for $N=2$, the system \eqref{DYN_WKH} is a planar competitive system,
\item for $N=3$, and for each $j\in [3]$, the system \eqref{DIFFDYN} in $\Sigma_j$ is cooperative, and every trajectory of \eqref{DYN_WKH} must eventually enter one of the $S_j$, as proved above.
\end{itemize}
Let now $N$ be arbitrary and $a_i = a$ for all $i \in [N]$. Then, the differential equation \eqref{DYN_WKH} is a gradient system, namely we have $F_i(\mathbf{J})= -\partial_{J_i} V(\mathbf{J})$ where the potential $V$ is given by 
\[
V(\mathbf{J})=\frac{\gamma}{2}\sum_{i=1}^NJ_i^2-a\log \sum_{i=1}^Ne^{J_i}.
\]
The conclusion then follows from the theory of gradient dynamical systems. 

Finally, notice that a necessary and sufficient condition for a first order differential equation $\dot{x} = F(x)$ to be a gradient system is that the vector field $F$ satisfies 
\[
\frac{\partial F_i}{\partial J_j} = \frac{\partial F_j}{\partial J_i},\ \forall(i,j) \in [N]^2.
\]
Clearly, in the heterogeneous case of distinct attractiveness, namely if $a_1\leq \cdots \leq a_N$ and $a_i \neq a_j$ for some  $i \neq j$, then the equation \eqref{DYN_WKH} is not a gradient system. 

\paragraph{Proof of statement {\em (ii)}.}
While Theorem \ref{MAINRES1} implies the existence of at least one stationary point for the system \eqref{DYN_WKH}, we prove this fact independently here. By Lemma \ref{EXIST}, any stationary point must belong to the set 
\[
\left\{\mathbf J\in (\R_\ast^+)^N\ :\ J_i\leq \frac{a_i}{\gamma},\ \forall i\in\{1,\cdots ,N\}\right\}.
\]
The existence of stationary point in this set follows from the Poincar\'e-Miranda Theorem (see e.g.\ \cite{Z86}) given that we have
\[
F_i(0)>0\quad\text{and}\quad F_i\left(\frac{a_i}{\gamma}\right)=a_i \left(\frac{e^{ J_i}}{\sum_{j=1}^{N} e^{J_j}}-1\right)<0,\ \forall i\in\{1,\cdots ,N\}.
\]

Now, let $F'(\mathbf J)$ be the derivative (Jacobian matrix) of $F$ at $\mathbf J$. We locate the spectrum of $F'(\mathbf J)+F'(\mathbf J)^T$ using the Gershgoring disk Theorem. The expression of the partial derivatives $\frac{\partial F_i}{\partial J_j}$ implies that the Gershgorin disks are included in the half-planes defined by 
\[
\mathrm{Re} (z)\leq -2\gamma+3 a_i\frac{e^{J_i}\sum_{k \neq i}e^{J_k}}{(\sum_{k=1}^{N}e^{J_k})^2}+\frac{e^{J_i}\sum_{k \neq i}a_ke^{J_k}}{(\sum_{k=1}^{N}e^{J_k})^2}\leq -2\gamma+4a_N\frac{e^{J_i}\sum_{k \neq i}e^{J_k}}{(\sum_{k=1}^{N}e^{J_k})^2}
\]

Clearly, the inequality $\frac{ab}{(a+b)^2}<\frac14$ implies that, when $\gamma>\frac{a_N}2$, the RHS of this inequality is negative, and hence that all trajectories of the equation \eqref{DYN_WKH} approach each other as $t\to +\infty$, see e.g.\ \cite{LS98}. When combined with the existence of the stationary point above, this proves that all trajectories must asymptotically converge to a unique stationary point.

\section{Analysis of the stationary points}\label{S-STATPT}
In this section, we focus on the analysis of stationary points and their dependence on parameters, especially in the domain $\gamma\leq \frac{a_N}2$. We do not aim to cover all cases, but rather illustrate the diversity of the phenomenology, especially as the ordering of preferences and the (non-)monotonicity in the friction parameter are concerned. 

In addition, notice that the results obtained for specific attractiveness ($N>2$), namely any linearly stable (or unstable) stationary point, can be extended to more general ones by continuation methods, namely by using the Implicit Function Theorem. This is a standard procedure that applies to any hyperbolic stationary point in a $C^k$-dynamical system. Actually, as in \cite{MS95}, by computing the norm of the resolvent of the Jacobian matrix at the stationary point, one can even estimate the size of those perturbations for which the point under consideration persists. 
\medskip

\noindent
{\bf Warning:} Throughout the section, we investigate the stationary point of the equation \eqref{DIFFDYN} for $j=N$. We shall drop the dependence on this index for the sake of notation.

\begin{figure}[ht]
\begin{center}
\includegraphics*[width=70mm]{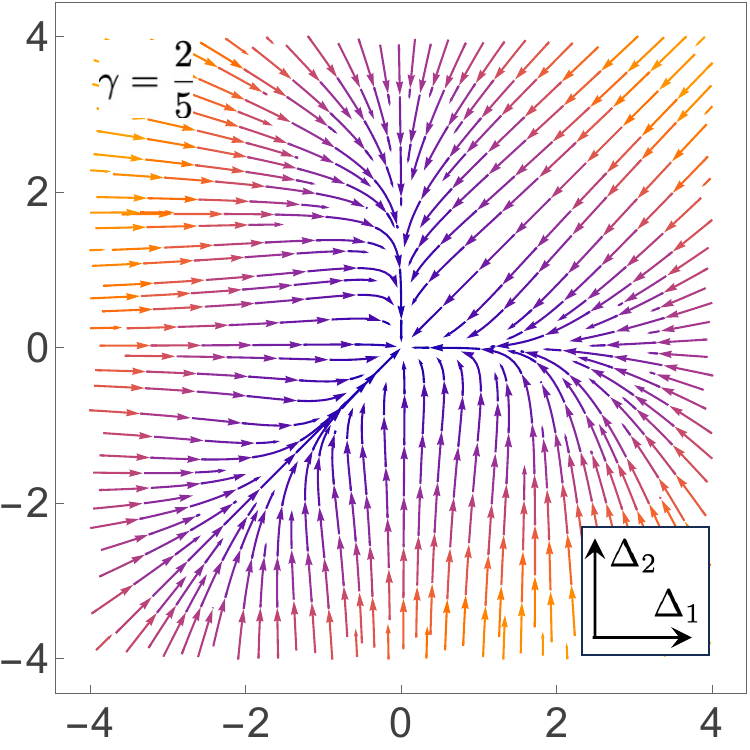}
\hspace{1cm}
\includegraphics*[width=70mm]{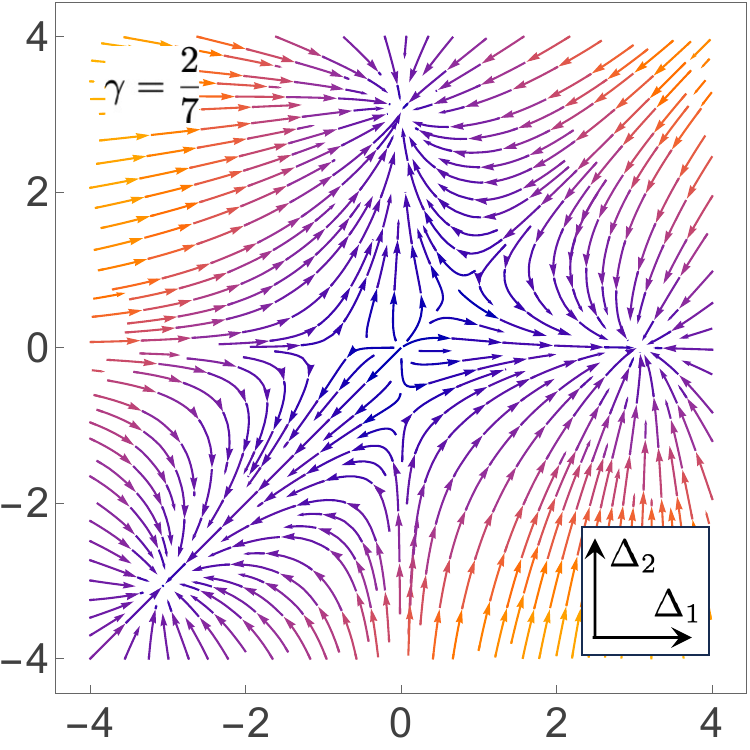}
\end{center}
\caption{Illustration of the results of Proposition \ref{FIX_HOMOG}. Stream lines of the vector field $G$ of the system \eqref{DIFFDYN} for $N=3$ and $a_1=a_2=a_3 = 1$. {\sl Left.}\ $\gamma=\frac25>\frac13$, there is a unique stationary point of $G$ (and hence of $F$) with coordinates $\Delta_1=\Delta_2=0$. {\sl Right.}\ $\gamma=\frac27<\frac13$, there are exactly 7 stationary points. Three points are stable. Following the Proposition (according to which, in the original variable $\mathbf J$, for each point, two coordinates are equal and the remaining one is larger), these points are respectively on the axis $\Delta_1=0$, on the axis $\Delta_2=0$ and on the diagonal $\Delta_1=\Delta_2$.}
\label{HOMOG_N=3}
\end{figure}

\paragraph{The homogeneous case of identical attractiveness.}
\begin{Thm}
\noindent
Assume that $a_i = a$ for all $i \in [N]$ for some arbitrary $a \in \R_+^\ast$. Then, the following statements hold.
\begin{itemize} 
\item[(i)] If $\gamma \geq \frac{a}{N}$,  then equation \eqref{DYN_WKH} has a unique stationary point (which must be globally attracting). By the permutation symmetry, the coordinates $J_i$ of this point must all be equal.
\item[(ii)] If $\gamma < \frac{a}{N} $, then equation \eqref{DYN_WKH} has $2^N-1$ stationary points. Among these points there are exactly $N$ points that are (asymptotically) stable\footnote{Throughout, 'stable' means 'asymptotically stable'.}. Such stable points have exactly $N-1$ equal coordinates and the remaining coordinate is larger than the other ones. 
\end{itemize}
\label{FIX_HOMOG}
\end{Thm}
Evidently, when $N > 2$, the domain of $\gamma$ for which there exists a unique stationary point is larger than in the case of an arbitrary $\mathbf{a}$ (Proposition \ref{CONTRACT}). Independently, the statement shows that when the friction becomes small, the permutation symmetry is broken and the dynamics spontaneously selects one seller among the others (the one for which the fixed point coordinate is the largest); the choice of whom depends on the initial condition. In other words, while the sellers are intrinsically indistinguishable, any initially slight asymmetry in the preferences is amplified in time and eventually yields a significantly larger preference for one seller over the others. An illustration of this result is given in Fig. \ref{HOMOG_N=3}.
\begin{proof}
The proof relies on the following statement.
\begin{Claim}
Assume that $a_i = a$ for all $i \in [N]$ for some arbitrary $a \in \R_+^\ast$ and let $\gamma \in \R^+_\ast$ be arbitrary. For every stationary point $\mathbf{J}$ of the equation \eqref{DYN_WKH}, there exists $E \subset [N]$ and $J_E, J_{E^c} \in \R_\ast^+$ such that 
\[
J_i=\left\{\begin{array}{cc} 
J_E \quad \forall i \in E\\
J_{E^c} \quad \forall i \in E^c
\end{array}\right.
\]
\end{Claim}
\begin{proof}
The vector $\mathbf \Delta$ associated with a stationary point $\mathbf J$ must be itself a stationary point of the equation \eqref{DIFFDYN}. 
When $a_i=a$ for all $i$, we have $G_i(0)=0$ for every $i \in [N-1]$, hence the homogeneous fixed point with all coordinates given by $J_i=\frac{a}{N\gamma}$, is always a stationary point in the homogeneous case of identical attractiveness. 

Let now $\mathbf{\Delta}$ be a stationary point of the equation \eqref{DIFFDYN} such that $\Delta_i \neq 0$ for some $i \in [N-1]$. The definition of $G$ implies that, for those coordinates $i$ such that $\Delta_i \neq 0$, we must have  
\[
\frac{e^{ \Delta_i}-1}{\Delta_i}=\frac{\gamma}{a}\left(\sum_{j=1}^{N-1} e^{ \Delta_j}+1\right)
\]
which gives 
\[\frac{e^{ \Delta_i}-1}{\Delta_i}=\frac{e^{ \Delta_j}-1}{\Delta_j},\]
for all $i,j$ such that $\Delta_{i},\Delta_{j} \neq 0$. The conclusion then follows from the fact the map $ \Delta \mapsto \frac{e^{\Delta}-1}{\Delta}$ is one-to-one on $\R_\ast$.  
\end{proof}

Let $\mathbf{\Delta}$ be a stationary point of the equation \eqref{DIFFDYN}. The Claim implies the existence of $\Delta \in \R$ such that for every $i\in \{1,\cdots ,N-1\}$ we have either $\Delta_i = 0$ or $\Delta_i = \Delta$. Moreover, letting $k = \# \{i \in \{1,\cdots ,N-1\}\ :\ \Delta_i \neq 0\}$, the quantity $\Delta$ must satisfy the following fixed point equation 
\begin{equation}
\Delta=g_k(\Delta)\quad\text{where}\quad g_k(\Delta)=\frac{a}{\gamma}\frac{e^{ \Delta}-1}{ke^{ \Delta}+N-k}.
\label{SOL_STAT}
\end{equation}
Easy computations show that the map $g_k$ has the following properties:
\begin{itemize}
\item it is increasing and we have $g_k(0)=0$, $g'_k(0)=\frac{a}{N\gamma}$ and 
\[
g''_k(\Delta)\left\{\begin{array}{ccl}
>0&\text{if}&\Delta <\log\frac{N-k}k\\
<0&\text{if}&\Delta >\log\frac{N-k}k
\end{array}\right.
\]
\item we have $\lim\limits_{\Delta \to +\infty} g_k(\Delta) = \frac{1}{k}$ and $\lim\limits_{\Delta \to -\infty} g_k(\Delta) = \frac{-1}{N-k}$.
\end{itemize}
Consequently, if $\gamma \geq \frac{a}{N}$, then for every $k\in\{1,\cdots ,N-1\}$, the point 0 is the unique fixed point of $g_k$. Indeed, the properties of first and second derivatives of $g_k$ imply that on one hand, no fixed point can exist in $\R_\ast^-$, and on the other hand, given also that $g'_k\left(\log\frac{N-k}k\right)=\frac{aN}{4\gamma k(N-k)}\leq 1$, no fixed point can exist in $\R_\ast^+$. It follows that equation \eqref{DYN_WKH} has a unique stationary point $J_i =\frac{a}{\gamma N}$ in this case, that must be globally attracting because this equation is a gradient system. Statement {\em (i)} is proved.

Consider now the case $\gamma<\frac{a}{N}$. The properties of $g_k$ above imply that, for every $k\in\{1,\cdots ,N-1\}$, the fixed point equation \eqref{SOL_STAT} has a unique solution in $\R_\ast^-$ and a unique solution in $\R_\ast^+$. Together with 0, this implies that the original equation \eqref{DYN_WKH} has exactly $1+2{\displaystyle\sum_{k=1}^{N-1}{N-1\choose k} }= 2^{N}-1$ stationary points. 

It remains to address the stability of the stationary points. To that goal, notice that the fact that that map $F$ commutes with the permutations of the $\mathbf{J}$-coordinates imply that $G$ commutes with the following transformations of $\boldsymbol{\Delta}$ 
\begin{itemize}
\item the transpositions of two coordinates $\Delta_i$ (corresponding to the transpositions of two $J_i$ with $i\neq N$),
\item for every $ j\in \{1,\cdots,N-1\}$, the transformation $\boldsymbol{\Delta}\mapsto \boldsymbol{\Delta}'$ defined by
\[
\Delta'_i=\left\{\begin{array}{cc}\Delta_i-\Delta_i&\text{if}\ i\neq j\\
-\Delta_j&\text{if}\ i=j
\end{array}\right.\quad \text{for}\ i\in \{1,\cdots,N-1\}
\]
(and corresponding to the transposition that exchanges $J_N$ with $J_j$)
\end{itemize}
In addition, it will be convenient to consider the following $N$ sectors of $(\R_\ast)^{N-1}$
\begin{itemize}
\item $S_0=(\R_\ast^-)^{N-1}$,
\item $S_j=\left\{\boldsymbol\Delta\in (\R_\ast)^{N-1}\ : \Delta_j=\max_i\Delta_i> 0\right\}$ for every $j\in\{1,\cdots ,N-1\}$.
\end{itemize}
Clearly, the union ${\displaystyle \bigcup_{j=0}^{N-1}}S_j$ covers $(\R_\ast)^{N-1}$ (and hence, it covers $\R^{N-1}$ modulo a set of zero Lebesgue measure), because given any $\boldsymbol{\Delta}\in (\R_\ast)^{N-1}$, either $\Delta_i< 0$ for all $i\in\{1,\cdots ,N-1\}$, or there exists $i\in\{1,\cdots ,N-1\}$ such that $\Delta_i> 0$. In the latter case, take the maximum of those positive coordinates.

Each such sector is forward invariant by one of the transformations above and is mapped onto the other ones by the other transformations. In order to determine those stationary points that are stable, it suffices to describe the dynamics in one of this sector, for instance in $S_0$.

Statement {\sl (ii)} then readily follows from the following result.
\begin{Claim}
Let $ \gamma < \frac{a}{N}$. The stationary point whose coordinates are given by 
\[
\Delta_i=\Delta_{N-1}^-,\ \forall i\in \{1,\cdots,N-1\},
\]
where $\Delta_{N-1}^-$ is the negative solution of the equation \eqref{SOL_STAT} for $k=N-1$, attracts every trajectory issued from an initial condition in $S_0$. 
\end{Claim}
\noindent
{\sl Proof.} Given $\boldsymbol\Delta\in \R^{N-1}$, let $\boldsymbol\Delta_1^c=\{\Delta_i\}_{i=2}^{N-1}\in \R^{N-2}$. We claim that for every $\boldsymbol\Delta_1^c\in (\R_\ast^-)^{N-2}$, there exists a unique $\Delta_1(\boldsymbol\Delta_1^c)<0$ such that 
\[
G_1(\boldsymbol\Delta)\left\{\begin{array}{ccl}
<0&\text{if}&\Delta_1\in (\Delta_1(\boldsymbol\Delta_1^c),0)\\
>0&\text{if}&\Delta_1\in (-\infty,\Delta_1(\boldsymbol\Delta_1^c))
\end{array}\right.
\]
This property is a consequence of the following features. 
\begin{itemize}
\item $G_1(\boldsymbol\Delta)=0$ whenever $\Delta_1=0$. 
\item Given $\boldsymbol\Delta_1^c\in \R^{N-2}$, we have ${\displaystyle\lim_{\Delta_1\to -\infty}} G_1(\boldsymbol\Delta)=+\infty$.
\item Assume that $\gamma<\frac{a}{N}$. Then, given $\boldsymbol\Delta_1^c\in (\R_\ast^-)^{N-2}$, we have $\frac{\partial G_1}{\partial \Delta_1}(\boldsymbol\Delta)>0$ when $\Delta_1=0$. 
\item Given $\boldsymbol\Delta_1^c\in (\R_\ast^-)^{N-2}$, the map $\Delta_1\mapsto G_1(\boldsymbol\Delta)$ is convex on $\R^-$. 
\end{itemize}
As for the dynamics of the equation \eqref{DIFFDYN}, the property above implies that every trajectory that stays forever in $S_0$ must approach the nullcline $\boldsymbol\Delta_1^c\mapsto \Delta_1(\boldsymbol\Delta_1^c)<0$. Moreover, the fact that the dynamics commutes with the transposition of the coordinates $\Delta_i$ implies that every trajectory forever in $S_0$ must in fact approach every nullcline $\boldsymbol\Delta_i^c\mapsto \Delta_i(\boldsymbol\Delta_i^c)<0$ for $i\in\{1,\cdots N-1\}$, where $\boldsymbol\Delta_i^c\{\Delta_j\}_{j\in\{1,\cdots ,N-1\},\ j\neq i}$ and $\Delta_i(\boldsymbol\Delta_i^c)$ is defined by applying the transposition $1\leftrightarrow i$ to $\Delta_1(\boldsymbol\Delta_1^c)$. It results that for every initial condition $\boldsymbol\Delta\in S_0$, the subsequent trajectory must stay in this sector forever and it must converge to the intersection of the nullclines, that is to say, to the unique stationary point in this sector.  
\end{proof}

\paragraph{Markets composed by two sellers ($N=2$).}
In order to get insights into the dynamics for arbitrary $\gamma$ and $\mathbf a$, we now focus on the case $N=2$. In this case, the dynamics in the segment $S = \{\mathbf{J} \in (\R_\ast^+)^2\ :\  \frac{J_1}{a_1}+ \frac{J_2}{a_2}= \frac{1}{\gamma}\}$ reduces to the following one differential equation for the variable $\Delta_1$
\begin{equation}
\dot\Delta_1 = G_1(\Delta_1) \quad \textrm{where}\quad G_1(\Delta_1)= -\gamma\Delta_1 + \frac{a_1 e^{\Delta_1} - a_2}{1+e^{\Delta_1}}.
\label{WKH_N=2}
\end{equation}
Assuming w.l.o.g.\ that $a_1 < a_2$, a complete characterization of the solutions of this differential equation is given in the next statement. 
\begin{Pro}
Assume that $a_1 < a_2$ are given. There exists $\gamma(a_1,a_2)\in \left(0,\frac{a_1+a_2}{4}\right)$ such that the following statements hold
\begin{itemize}
\item[(i)] If $\gamma > \gamma(a_1,a_2)$, then the equation \eqref{WKH_N=2} has a unique globally attracting stationary point, whose coordinate is negative.
\item [(ii)] If $\gamma=\gamma(a_1,a_2)$, then the equation \eqref{WKH_N=2} has two stationary points. The smallest one has negative coordinate and is stable. The largest one has positive coordinate and is semi-stable (unstable wrt.\ negative perturbations, stable with respect to positive ones).
\item[(iii)] If $\gamma < \gamma(a_1,a_2)$, then the equation \eqref{WKH_N=2} has three stationary points. The smallest and the largest ones, which have negative and positive coordinate respectively, are locally asymptotically stable. The remaining one is unstable.
\end{itemize}
\label{N=2_HETEROG}
\end{Pro}
As before, this statement implies that when $\gamma > \gamma(a_1,a_2)$, in every trajectory, we must have $J_1(t) < J_2(t)$ for $t$ sufficiently large, {\sl viz.}\ preferences must be eventually ordered as the attractiveness are. 

The situation is more involved at weak friction. The bifurcation at $\gamma = \gamma(a_1,a_2)$ that generates a positive stable stationary point when the friction decreases below that threshold shows that,  depending on the trajectory, the eventual ordering of the preferences may be opposite to the one of the attractiveness ($J_1(t) > J_2(t)$ for $t$ sufficiently large even though $a_1 < a_2$). As paradoxical as this may appear, and as already observed in \cite{WKH00}, the seller with the largest attractiveness needs not be the one with the largest preference in the long term.
\begin{proof}
The derivative of $G_1$ writes
\[
G'_1(\Delta_1)= -\gamma + \frac{(a_1+a_2) e^{\Delta_1} }{(1+e^{\Delta_1})^2}
\]
It is easy to verify that the second term in this expression is a positive, even and unimodal function of $\Delta_1$, whose maximum at $\Delta_1=0$ is equal to $\frac{a_1+a_2}4$. Therefore, when $\gamma\geq \frac{a_1+a_2}4$, the map $G_1$ is decreasing on $\R$ and thus, since $G_1(\mp\infty)=\pm\infty$, it has a unique zero, which then must be a globally attracting stationary point of \eqref{WKH_N=2}. This proves statement {\em (i)}.

When $\gamma\in (0, \frac{a_1+a_2}4)$, let $\Delta_1^\pm=\log\left(\frac{a_1+a_2}{2\gamma}-1\pm\sqrt{\frac{a_1+a_2}{\gamma}\left(\frac{a_1+a_2}{4\gamma}-1\right)}\right)\in \R$ be such that $e^{\Delta_1^\pm}$ are the two real roots of the equation
\[
(1+x)^2=\frac{a_1+a_2}{\gamma}x.
\]
Notice that $\Delta_1^-=-\Delta_1^+<0$. Moreover, the map $G_1$ is decreasing on $(-\infty,\Delta_1^-)\cup (\Delta_1^+,+\infty)$ and it is increasing on $(\Delta_1^-,\Delta_1^+)$, and we still have $G_1(\mp\infty)=\pm\infty$. 

In order to determine the number of zeros of $G_1$, we then evaluate the location of $G_1(\Delta_1^\pm(\gamma))$ (adding the explicit dependence on $\gamma$), with respect to 0 as $\gamma$ varies in the interval $(0, \frac{a_1+a_2}4)$. One easily checks that the following properties holds.
\begin{itemize}
\item The map $\gamma \mapsto G_1(\Delta_1^-(\gamma))$ is increasing on $(0, \frac{a_1+a_2}4)$ and $\gamma \mapsto G_1(\Delta_1^+(\gamma))$ is decreasing,
\item $G_1(\Delta_1^-(\frac{a_1+a_2}4))=\frac{a_1-a_2}2<0$.
\item $G_1(\Delta_1^+(0^+)) = a_1>0$ and $G_1(\Delta_1^+(\frac{a_1+a_2}4)) = \frac{a_1-a_2}{2}$.
\end{itemize}
Therefore, there exists $\gamma(a_1,a_2)\in \left(0,\frac{a_1+a_2}{4}\right)$ such that $G_1(\Delta_1^+(\gamma))>0$ for all $\gamma\in (0,\gamma(a_1,a_2))$ and $G_1(\Delta_1^+(\gamma))<0$ for all $\gamma\in (\gamma(a_1,a_2),\frac{a_1+a_2}{4})$. Moreover, we have $G_1(\Delta_1^-(\gamma))<0$ for all $\gamma\in \left(0,\frac{a_1+a_2}{4}\right)$. The Proposition, and the existence and stability of the stationary points, then immediately follow.
\end{proof}

\paragraph{Two-cluster dynamics in two-cluster markets.} 
A direct extension of the previous studies considers the case of a markets with attractiveness of only two kinds, in which the preferences are synchronized accordingly. While this is a very specific case, it shows that the phenomenological changes may not be monotonic with the friction parameter (see statement {\em (ii)} in the Proposition below). 

To see this, let $N>2$ be arbitrary, assume the existence of $k\in [N-1]$ such that (w.l.o.g.) we have
\[
a_1=\cdots = a_k<a_{k+1}=\cdots =a_N,
\]
and consider the two clusters configurations in $S$ such that 
\[
J_1 =\cdots = J_k\quad \text{and}\quad J_{k+1}=\cdots =J_N.
\]
As before, the dynamics in this set is governed again by the one-dimensional differential equation $\dot\Delta_1=G_1(\Delta_1)$ where $G_1$ now writes
\[
G_1(\Delta_1)=-\gamma\Delta_1 + \frac{a_1 e^{\Delta_1} - a_N}{N-k+ke^{\Delta_1}}.
\]
As announced, the phenomenology depends on $k,N$ and on the values of $a_1$ and $a_N$. Let
\[
A(a_1,a_N,k,N)=(N-k)a_1\left(1 -\frac12\log\frac{N-k}{k}\right)-k a_N \left(1 +\frac12\log\frac{N-k}{k}\right).
\]
\begin{Pro}
(i) If $k=N-k$ or if $k>N-k$ and $A(a_1,a_N,k,N)\leq 0$, then a similar conclusion as in Proposition \ref{N=2_HETEROG} applies, with $\gamma(a_1,a_2)$ replaced by some $\gamma(a_1,a_N,k,N)\in (0,\frac{a_1}{4k}+\frac{a_N}{4(N-k)})$. 

\noindent
(ii) If $k>N-k$ and $A(a_1,a_N,k,N)>0$ (which is compatible with $a_1<a_N$), then there exist $\gamma_1<\gamma_2<\gamma_3\in (0,\frac{a_1}{4k}+\frac{a_N}{4(N-k)})$ such that 
\begin{itemize}
\item if $\gamma \in (\gamma_1,\gamma_2)$ or $\gamma>\gamma_3$, then the equation $\dot\Delta_1=G_1(\Delta_1)$ has a unique globally attracting stationary point, whose coordinate is negative,
\item if $\gamma\in (0,\gamma_1)\cup (\gamma_2,\gamma_3)$, then the equation $\dot\Delta_1=G_1(\Delta_1)$ has three stationary points. The smallest and the largest ones are locally asymptotically stable. The remaining one is unstable.
\end{itemize}
\end{Pro}
\begin{proof}
The derivative of $G_1$ writes
\[
G'_1(\Delta_1)= -\gamma + \frac{((N-k)a_1+ka_N) e^{\Delta_1} }{(N-k+ke^{\Delta_1})^2}
\]
It is easy to verify that the second term in this expression is a positive and unimodal function of $\Delta_1$, whose maximum at $\Delta_1=\log\frac{N-k}{k}$ is equal to $\frac{(N-k) a_1 +k a_N}{4k(N-k)}=\frac{a_1}{4k}+\frac{a_N}{4(N-k)}$. We also have $G_1(\pm\infty)=\mp\infty$. Hence, as in the case $N=2$, it follows that, when $\gamma\geq \frac{a_1}{4k}+\frac{a_N}{4(N-k)}$, the map $G_1$ has a unique zero, which then must be a globally attracting stationary point of the dynamics in the subspace $\Delta_1=\cdots =\Delta_{k}$ and $\Delta_{k+1}=\cdots =\Delta_{N-1}=0$.

When $\gamma\in (0, \frac{a_1}{4k}+\frac{a_N}{4(N-k)})$, let 
\[
\Delta_1^\pm=\log\frac{N-k}{k}+\log\left(\frac{(N-k) a_1 +k a_N}{2k(N-k)\gamma}-1\pm\sqrt{\frac{(N-k) a_1 +k a_N}{k(N-k)\gamma}\left(\frac{(N-k) a_1 +k a_N}{4k(N-k)\gamma}-1\right)}\right)\in \R,
\]be such that $e^{\Delta_1^\pm}$ are the two real roots of the equation
\[
(N-k+k x)^2=\frac{((N-k)a_1+ka_N)}{\gamma}x.
\]
Notice that 
\[
\Delta_1^--\log\frac{N-k}{k}=-\left(\Delta_1^+-\log\frac{N-k}{k}\right)=\log\left(\frac{b}2-1-\sqrt{b(\frac{b}4-1)}\right)<0,
\]
where $b=\frac{(N-k) a_1 +k a_N}{k(N-k)\gamma}\in (4,+\infty)$.  
Moreover, the map $G_1$ is decreasing on $(-\infty,\Delta_1^-)\cup (\Delta_1^+,+\infty)$ and it is increasing on $(\Delta_1^-,\Delta_1^+)$, and we still have $G_1(\mp\infty)=\pm\infty$. 

We aim at determining the sign of $G_1(\Delta_1^\pm(\gamma))$  as $\gamma$ varies in the interval $(0,  \frac{a_1}{4k}+\frac{a_N}{4(N-k)})$. Explicit calculations yield the following features
\begin{itemize}
\item The sign of the derivative of $\gamma\mapsto G_1(\Delta_1^\pm(\gamma))$ is the opposite of the sign of $\Delta_1^\pm(\gamma)$. 
\item If $N-k\leq k$, then $\Delta_1^-(\gamma)<0$ for all $\gamma$. Moreover 
\[
\Delta_1^+(\gamma)\left\{\begin{array}{ccl}
>0&\text{if}&\gamma \in (0,\frac{(N-k)a_1+k a_N}{N^2})\\
<0&\text{if}&\gamma \in (\frac{(N-k)a_1+k a_N}{N^2},\frac{(N-k)a_1+k a_N}{4k(N-k)})
\end{array}\right.
\]
Notice that the second case does not occur if $N-k=k$.
\item If $N-k\geq k$, then $\Delta_1^+(\gamma)>0$ for all $\gamma$. Moreover 
\[
\Delta_1^-(\gamma)\left\{\begin{array}{ccl}
<0&\text{if}&\gamma \in (0,\frac{(N-k)a_1+k a_N}{N^2})\\
>0&\text{if}&\gamma \in (\frac{(N-k)a_1+k a_N}{N^2},\frac{(N-k)a_1+k a_N}{4k(N-k)})
\end{array}\right.
\]
Notice that the second case does not occur if $N-k=k$.
\item We have $G_1(\Delta_1^\pm(\frac{(N-k)a_1+k a_N}{4k(N-k)}))=\frac{(N-k)a_1\left(1 -\frac12\log\frac{N-k}{k}\right)-k a_N \left(1 +\frac12\log\frac{N-k}{k}\right)}{2k(N-k)}$.
\item $G_1(\Delta_1^+(0^+))=\frac{a_1}{k}$ and $G_1(\Delta_1^-(0^+))=-\frac{a_N}{N-k}$.
\end{itemize}
One can decompose the analysis according to the following cases.
\begin{itemize}
\item If $N-k=k$ the situation is as for $N=2$. 
\item If $N-k<k$ and $(N-k)a_1\left(1 -\frac12\log\frac{N-k}{k}\right)-k a_N \left(1 +\frac12\log\frac{N-k}{k}\right)\leq 0$, then the situation is also as for $N=2$ because even though $G_1(\Delta_1^+)$ increasing for $\gamma \in (\frac{(N-k)a_1+k a_N}{N^2},\frac{(N-k)a_1+k a_N}{4k(N-k)})$, it cannot become positive in this interval since $G_1(\Delta_1^+(\frac{(N-k)a_1+k a_N}{4k(N-k)}))<0$. This completes the proof of statement {\em (i)}.
\item More interestingly is the case $N-k<k$ and $(N-k)a_1\left(1 -\frac12\log\frac{N-k}{k}\right)-k a_N \left(1 +\frac12\log\frac{N-k}{k}\right)> 0$ because then we have 
\[
G_1(\Delta_1^+(\frac{(N-k)a_1+k a_N}{N^2}))=-\frac{a_1+a_N}{N}<0
\]
Hence, given the monotonicity of the maps $\gamma\mapsto G_1(\Delta_1^\pm(\gamma))$, the interval $(0,\frac{(N-k)a_1+k a_N}{4k(N-k)})$ decomposes into 4 intervals $(0,\gamma_1), (\gamma_1,\gamma_2),(\gamma_2,\gamma_3)$ and $(\gamma_3,\frac{(N-k)a_1+k a_N}{4k(N-k)})$ that are such that 
\begin{itemize}
\item $G_1(\Delta_1^-(\gamma))<0<G_1(\Delta_1^+(\gamma))$ on $(0,\gamma_1)$ and on $(\gamma_2,\gamma_3)$ and there are three stationary points in this case.
\item $G_1(\Delta_1^+(\gamma))<0$ on $(\gamma_1,\gamma_2)$ and $G_1(\Delta_1^-(\gamma))>0$ on $(\gamma_3,\frac{(N-k)a_1+k a_N}{4k(N-k)})$ and the stationary point is unique in these two cases.
\end{itemize}
Statement {\em (ii)} immediately follows.
\end{itemize}
\end{proof}

\section{Conclusion}
In this paper, we have provided an extensive mathematical analysis of the dynamics of the model defined in \cite{WKH00} in the particular case where the buyers' characteristics are all equal. We have proved convergence to stationary points for almost every trajectory and we determined these points, their stability and their friction-dependent bifurcations in some illustrative examples of distributions of sellers' attractiveness. In particular, we showed that the occurrence of multiple stable points as the friction coefficient decreases may not be monotonous, {\sl ie.}\ the points may emerge at some value, then disappear and then emerge again. 

We conclude the paper by few considerations about perspectives and suggestions for future works. First, in the framework of network dynamical systems \cite{GS03}, it would be interesting to investigate, for heterogeneous attractivenesses, those symmetry-breaking bifurcations that generate stationary points outside synchrony subspaces. In particular, along the lines of Theorem \ref{FIX_HOMOG}, one way investigate those symmetries of the (stable) stationary points depending on the symmetry of the attractiveness. 

As modelling is concerned, to address heterogeneous buyers populations is of primary interest given the empirical evidences of various types of customers in wholesale fresh product markets \cite{C10,VE11}. Some buyers (the 'loyal' ones) are prone to return to the same providers, other (the 'nomads') do not have preferred providers and systematically visit all sellers before making their choice. Such heterogeneities could first be apprehended via a multi-population formalism in which attractiveness also depends on buyer's type. Yet, agent-based models can also be envisaged in order to address the dynamics at the individuals level.  

More elaborated modelling may also consider that attractiveness are not constant in time but depend on a number of factors, including externalities and feedback from the sellers. In particular, our papers \cite{EF24,EF25} consider some negative feedback and showed that it induces oscillatory behaviours that may or may not asymptotically vanish in the long term depending on parameters. Along these lines, it would be interesting to investigate in the framework of \cite{WKH00}, how feedback impacts the dynamics and the convergence to stationary points in particular.

\end{document}